\documentclass[9pt,shortpaper,twoside,web]{ieeecolor}
\usepackage{generic}
\usepackage{cite}
\usepackage{amsmath,amssymb,amsfonts}
\usepackage{algorithmic}
\usepackage{graphicx}
\usepackage{textcomp}
\usepackage{multirow}
\usepackage{booktabs}

\usepackage{epstopdf}

\def\BibTeX{{\rm B\kern-.05em{\sc i\kern-.025em b}\kern-.08em
    T\kern-.1667em\lower.7ex\hbox{E}\kern-.125emX}}
\begin{document}
\title{Alzheimer's Disease Prediction via Brain Structural-Functional Deep Fusing Network}
\author{Qiankun Zuo, Junren Pan, and Shuqiang Wang
\thanks{Corresponding author: Shuqiang Wang (email: sq.wang@siat.ac.cn)}
\thanks{Qiankun Zuo, Junren Pan and Shuqiang Wang are with the Shenzhen Institutes of Advanced Technology, Chinese Academy of Sciences, Shenzhen, 518055, China}}

\maketitle

\begin{abstract}
Fusing structural-functional images of the brain has shown great potential to analyze the deterioration of Alzheimer's disease (AD). However, it is a big challenge to effectively fuse the correlated and complementary information from multimodal neuroimages. In this paper, a novel model termed cross-modal transformer generative adversarial network (CT-GAN) is proposed to effectively fuse the functional and structural information contained in functional magnetic resonance imaging (fMRI) and diffusion tensor imaging (DTI). The CT-GAN can learn topological features and generate multimodal connectivity from multimodal imaging data in an efficient end-to-end manner. Moreover, the swapping bi-attention mechanism is designed to gradually align common features and effectively enhance the complementary features between modalities. By analyzing the generated connectivity features, the proposed model can identify AD-related brain connections. Evaluations on the public ADNI dataset show that the proposed CT-GAN can dramatically improve prediction performance and detect AD-related brain regions effectively. The proposed model also provides new insights for detecting AD-related abnormal neural circuits.
\end{abstract}

\begin{IEEEkeywords}
Multimodal fusion, brain network computing, swapping bi-attention mechanism, generative adversarial strategy.
\end{IEEEkeywords}

\section{Introduction}
\label{sec:introduction}

Alzheimer's disease (AD) is a progressive and irreversible neurodegenerative disease that has become the primary cause of dementia among the elderly~\cite{dadar2017validation}. According to statistics\cite{AS}, there are more than fifty million AD sufferers all around the world. As the patients progress towards AD, they will lose cognitive abilities, including the ability to remember or think, and finally lose the ability to care for themselves~\cite{scheltens2016alzheimer,Association2018,colom2020clinical}. The high prevalence of AD creates a heavy financial strain on governments as well as the patients' families. Thanks to the development of artificial intelligence, researchers are studying and analyzing AD by using machine learning-based tools~\cite{mci2,mci3,zong2022multiscale}. Yet, the exact cause of AD is still unknown. One of the primary causes of the aforementioned challenges is that the brain is a highly complex system, and carrying out cognitive tasks needs topological communications between regions of interest (ROIs). Therefore, the study of brain network computation is beneficial to the diagnosis and analysis of cognitive brain diseases, as well as to explore potential biomarkers.

The graph is usually used to define the brain network. The ROIs of the whole brain network represent the graph's nodes. The relationships between the ROIs in a brain network represent their edges\cite{zhang2019strength,song2022multi}. In the brain network, there are two main types of connectivities, including functional connectivity (FC) and structural connectivity (SC). A definition of FC is the relationship between two ROIs' bioelectric signals, which can be observed through functional magnetic resonance imaging (fMRI). The neural fiber connection strength between brain regions is defined as SC. It makes use of diffusion tensor imaging (DTI) to measure water molecular dispersion motion.

Many studies have employed either FC or SC to discover certain AD-related characteristics that are not detectable using traditional imaging techniques \cite{JeonKang2020,sh2}. They demonstrate that brain network approaches in AD studies have more benefits than the traditional imaging approach. Colclough et al.\cite{GroupSparse2} presented a hierarchical inverse covariance algorithm to simultaneously infer connectivity strength at both subject and population levels. To preserve the brain's local geometry or manifold structure, Yu et al. \cite{GroupSparse3} applied a weighted graph regularized sparse representation (WGraphSR) method to obtain brain connectivity. It can not only boost the MCI prediction performance but also reveal more valuable connections associated with MCI. To explore the causal relations between ROIs and lower individual differences, Li et al. \cite{GroupSparse1} proposed a sparse constrained connectivity inference model to construct a functional network from functional time series and then built a multilayer perceptron classifier for MCI detection. Besides, the high-order features can be represented by a hypergraph, where multiple ROIs share the same edge. They \cite{dynamic6} proposed a hypergraph learning-based method to construct brain functional connectivity for a better understanding of the brain's overall structure. To fully make use of the dynamic interactions among brain regions, \cite{dynamic5} first derived multiple low-order functional connectivity networks (FCNs) from a series of sliding windows and then constructed a high-order FCN by measuring the topographical similarity between FCNs. The disease-related biomarkers can be successfully recognized. However, most existing investigations of brain networks concentrate on single-modal imaging, which makes it impossible to focus on the integration of structural and functional connectivity information.

Single-modal imaging would disregard the opportunity to use complementary cross-modal information to more deeply understand AD since it might only partially contain AD-related information. Therefore, in the computation of medical imaging, multimodal brain networks\cite{ZhangShen2012,multi1,multi2,multi3} are becoming more prevalent. As multimodal brain networks are heterogeneous and concealed in various types of neuroimaging data, how to properly exploit complementary information between modalities is crucial for structural-functional deep fusion. The majority of current methods only employ linear interactions to fuse structural and functional information\cite{abrol2019multimodal}. For example, ~\cite{plis2018reading} proposed a novel deep neural network-based model to effectively fuse structural MRI and functional MRI by finding linkages between bimodal images. It reveals a significant correlation between the impairments in schizophrenia and the function/structure alignment score. The work in ~\cite{dekhil2019personalized} developed a computer-aided detection system to combine structural and functional abnormalities for autism prediction, which discovered autism-related areas affected by impairment loss. Similarly, they ~\cite{hirjak2020multimodal} utilized multimodal magnetic resonance imaging to study the abnormal structure-function patterns in catatonia. The co-altered interactions in the brain are founded to facilitate visuospatial functions and motor behavior. Due to the fact that changes in brain structure and functional connectivity cannot be entirely explored by linear correlations~\cite{zhang2021deep}, we applied a graph convolutional network (GCN)-based network to deeply fuse structural and functional connectivity information for mild cognitive impairment diagnosis. The topological properties are fully explored by iteratively updating the fused deep connectome. However, previous studies\cite{Honey,Li15,zuo2021multimodal} showed that strong FC typically follows strong SC but rarely the other way around. Clinical research \cite{Da2015,LeiCheng2020,Cao2018} demonstrates that certain regions can compensate for a lowered SC when it happens by increasing functional activity between RIOs.

GANs can be thought of as generative models based on variational inference. It has been demonstrated that GANs are an effective model to learn multivariate distributions\cite{GAN}. GANs are currently employed successfully in several areas of medical image analysis\cite{shgan2,shgan5,shgan6}. Transformers \cite{trans1} have demonstrated their strong capacity for sequential analysis and successful applications in natural language processing (NLP). The key point is the self-attention mechanism's function in identifying nonlinear connections between inputs. Transformers have recently been used for image tasks as a result of their successful NLP applications~\cite{trans2,trans3,trans4}. However, transformers haven't been thoroughly studied in the context of brain networks. Therefore, the CT-GAN is proposed in this study to combine structural and functional connectivities for AD analysis. Using a swapping bi-attention mechanism, the proposed cross-modal transformer-based network generates multimodal connectivity (MC). A dual-channel separator and a generative adversarial strategy are used to optimize the training of the CT-GAN to maintain the learned MC's robustness. The main contributions to this work are as follows:

\begin{itemize}
	\item The proposed CT-GAN is proposed to transform the fMRI and DTI into multimodal connectivity for AD analysis by combining the generative adversarial strategy. It not only learns the topological characteristics of non-Euclidean space but also deeply fuses the complementary information in an efficient end-to-end manner.
	\item The swapping bi-attention mechanism (SBM) is developed to effectively align functional information to microstructural information and enhance the complementary information between bimodal images.
	\item The dual-channel separator with cross-weighting scheme is devised to decompose the multimodal connectivity into functional and structural connectivities, which preserves the global topological information and ensures the high quality and diversity of the generated connectivities.
\end{itemize}

The remaining sections of this paper are divided into the following sections: Section~\ref{s2} presents the overall design of the proposed CT-GAN. The experimental findings, including generation evaluation and classification performance, are presented in Section~\ref{s3}. The primary remarks of this study are presented in Section~\ref{s4}.

\begin{figure*}[htbp]
	\centering
	\includegraphics[width=0.95\textwidth]{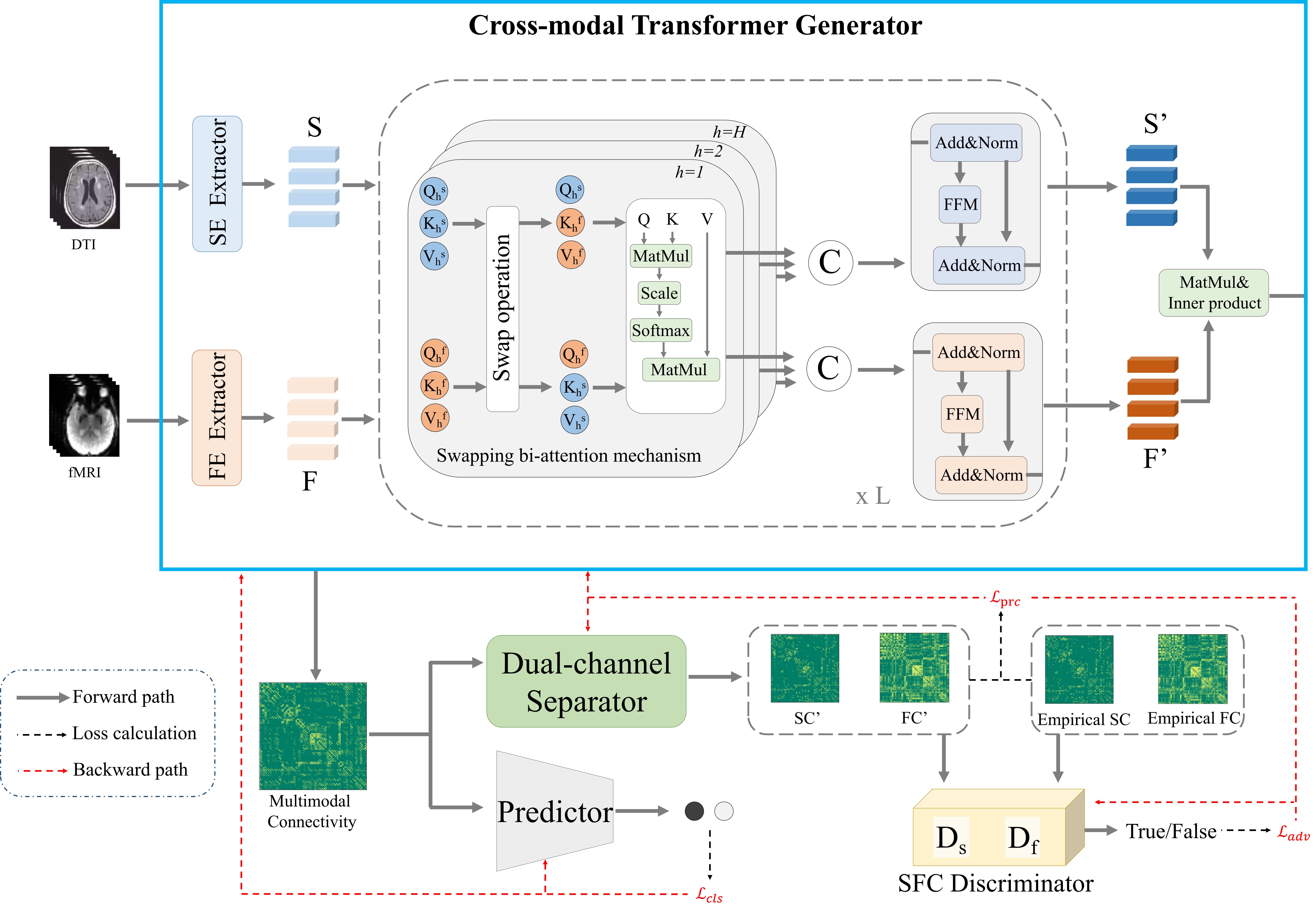}
	\caption{The framework of the proposed CT-GAN, including four parts: the cross-modal transformer generator, the dual-channel separator, the SFC discriminator, and the predictor. $S$ represents the structural embedding, and $F$ represents the functional embedding. The framework aims to generate multimodal connectivity from DTI and fMRI.
	} \label{fig1}
\end{figure*}

\section{Method}
\label{s2}
Fig.~\ref{fig1} presents the architecture of the proposed CT-GAN. Given bimodal images (i.e., fMRI and DTI), the proposed model can learn a non-linear learning network  that can transform imaging space to topological connectivity space. To obtain multimodal connectivity, the CT-GAN is comprised of four components: 1) the cross-modal transformer generator ($G$) that is used for inference and generates multimodal connectivity; 2) the dual-channel separator (DS) that decomposes the multimodal connectivity into SC and FC; 3) the structural-functional consistency (SFC) discriminator, which contains two sub-discriminators (i.e., $D_s$ and $D_f$). Each of them discriminates whether an SC or FC comes from the proposed generator or the software toolboxes; 4) the predictor that assigns AD stages based on the generated multimodal connectivity.

\subsection{Cross-modal Transformer Generator}

\subsubsection{Embedding Extractor}

To embed ROI-based features into the transformer network, a routine convolutional neural network is adopted to extract features from medical imaging. We designed two extractors to obtain rough ROI-based features, including a structural embedding (SE) extractor and a functional embedding (FE) extractor. As shown in the upper left of Fig.~\ref{fig1}, the embedding $S$ is computed by successive convolutional filters on the DTI. Specifically, we first design four down-sampling operations with a $3 \times 3$ kernel of a 2-step size of $2$ to extract local feature maps. The feature maps are then passed through $1 \times 1$ filters to fix the channel at $q$. Finally, each channel map is combined with the brain anatomical information ($x,y,z,v$)~\cite{zuo2022constructing} to align the features for every brain region. Similar operations are conducted on the fMRI. The output embeddings $\bold S$ and $\bold F$ are given below:

\begin{equation}
	\bold S = SE(DTI,x,y,z,v),\quad \bold F = FE(fMRI,x,y,z,v)
\end{equation}
where $\bold S \in \mathbb{R}^{N\times q}$ , $\bold F \in \mathbb{R}^{N\times q}$.

\subsubsection{Swapping Bi-attention Mechanism}

This study's main concept is to leverage transformers' bi-attention mechanism to merge structure-function data for fMRI and DTI. A transformer gradually projects an input embedding as $\bold S\in \mathbb{R}^{N\times q}$ to a target feature embedding as $\bold F\in \mathbb{R}^{N\times q}$, where $N$ represents the overall number of ROIs. The following is a description of the transformer learning process: In the beginning, a collection of query matrices $Q$, key matrices $K$, and value matrices $V$ are computed using a linear projection.

\begin{equation}
	Q = XW^q,\quad K = XW^k,\quad V = XW^v
\end{equation}
here, $X$ represents the $\bold S$ or $\bold F$. $W^q \in \mathbb{R}^{N\times q}$ , $W^k \in \mathbb{R}^{N\times q}$, and $W^v \in \mathbb{R}^{N\times q }$ are weight parameters. The attention of $Q$, $K$, and $V$ can be computed by the following formula:
\begin{equation}
	\text{Att}(Q,K,V) = \text{softmax}\Big(\frac{QK^T}{\sqrt{q}}\Big)V.
\end{equation}
in the proposed model, we design $H$ heads for each modality to focus on different parts of the learned embeddings. The tokens can be computed by

\begin{equation}
	Q_h^S = \bold SW_h^{qs},\quad K_h^S = \bold SW_h^{ks},\quad V_h^S = \bold SW_h^{vs}
\end{equation}
\begin{equation}
	Q_h^F = \bold FW_h^{qf},\quad K_h^F = \bold FW_h^{kf},\quad V_h^S = \bold FW_h^{vf}
\end{equation}
where, $h$ is the index of all $H$ heads. Each head has the dimension $q/H$. Each token (i.e., $Q_h^S,K_h^S,V_h^S$) has the same size, $N \times q/H$.

Then, we exchange the tokens between the two modalities and fuse the intermediate features adaptively. For structural modality, the token $Q_h^S$ is combined with the other two tokens ($K_h^F$ and $V_h^F$) to adaptively bring additional functional information into the structural features. And vice versa for functional modality. The structural and functional swapping bi-attention can be defined as

\begin{equation}
	S_h^{Att} = \text{F2SAtt}(Q_h^S,K_h^F,V_h^F) = \text{softmax}\Big(\frac{Q_h^S (K_h^F)^T}{\sqrt{q/H}}\Big)V_h^F.
\end{equation}
\begin{equation}
	F_h^{Att} = \text{S2FAtt}(Q_h^F,K_h^S,V_h^S) = \text{softmax}\Big(\frac{Q_h^F (K_h^S)^T}{\sqrt{q/H}}\Big)V_h^S.
\end{equation}

Finally, the output of the cross-modal swapping bi-attention can be formulated by
\begin{equation}
	S^{Att} = [S_1^{Att}, S_2^{Att}, ..., S_H^{Att}]
\end{equation}
\begin{equation}
	F^{Att} = [F_1^{Att}, F_2^{Att}, ..., F_H^{Att}]
\end{equation}
where, $[,]$ denotes the concatenation along the ROI feature direction.

\begin{figure*}[htbp]
	\centering
	\includegraphics[width=0.8\textwidth]{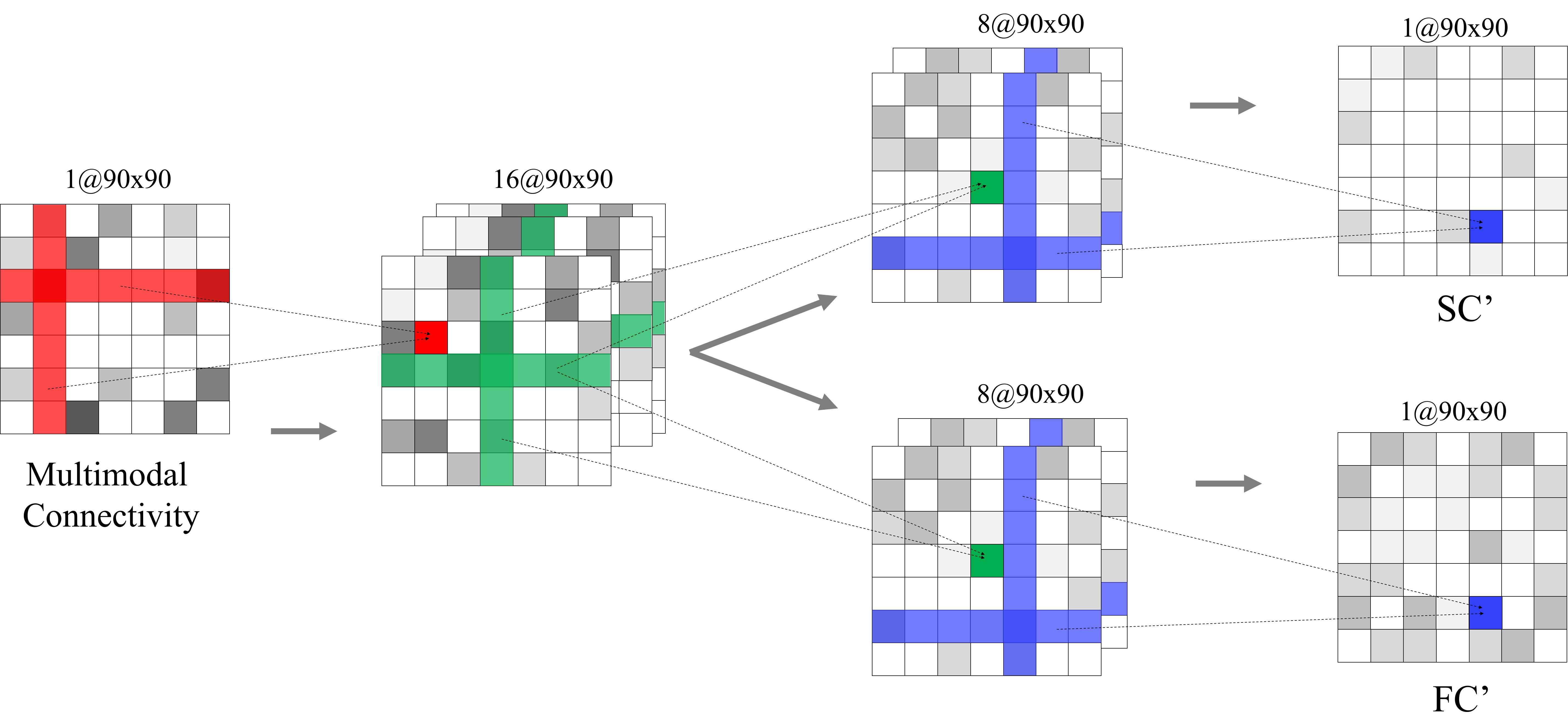}
	\caption{The network architecture of the dual-channel separator. Given the multimodal connectivity, it outputs the structural connectivity and functional connectivity.
	} \label{fig2}
\end{figure*}

\subsubsection{Feed Forward Mapping}

After the attention mechanism for each ROI feature, one fully connected layer (FC) is applied to adjust the attention ROI feature ($\bold S, \bold F$) into a feature sequence by adding and multiplying operations. Let's use the symbols $\bold F$ and $\bold S$ to represent the mixed functional feature sequence and mixed structural feature sequence, respectively. Recall that each ROI's functional feature sequence is updated by $\bold F' \in \mathbb {R}^{N \times q}$, and the updated structural feature is represented by $\bold S' \in \mathbb{R}^{N \times q}$, where $n$ is the sum of all the ROIs. The formula is defined as

\begin{equation}
	\bold S = \bold S + S^{Att}
\end{equation}
\begin{equation}
	\bold F = \bold F + F^{Att}
\end{equation}
\begin{equation}
	\bold S' = \bold S + FFM(\bold S)
\end{equation}
\begin{equation}
	\bold F' = \bold F + FFM(\bold F)
\end{equation}

\subsubsection{Connectivity Computation}

After $L$ layers of transformer, we obtained the mixed structural and functional features $\bold S'$ and $\bold F'$. These mixed features contain common and unique information for both modalities. We first project one modal feature onto the other modal feature, then compute the relationship of paired ROIs with the following formula:

\begin{equation}
	\text{MC} = \bold S' \bold F' \bold F'^T \bold S'^T.
\end{equation}
here, $MC$ is the final multimodal connectivity (MC) with the size $N \times N$.

\subsection{Dual-channel Separator}

The MC combines both structural and functional connectivity information. To stabilize the learning process, the SC and FC need to be recovered from MC. As shown in Fig.~\ref{fig2}, the dual-channel separator projects the MC back to connectivity in each modality domain. Considering the topological properties of the human brain, we adopt the cross-weighting scheme to extract global connectivity information for better detachment between structural and functional connectivity. It consists of two branches, which share the $1st$ layer and have different weighting parameters in the $2nd$ and $3rd$ layers, respectively. The filter is a cross-shape parameter with step size 1. The input and the output for each layer have the same size except for different channels. Finally, the $3rd$ layer outputs the reconstructed SC and FC.

\begin{figure}[h!]
	\centering
	\includegraphics[width=\columnwidth]{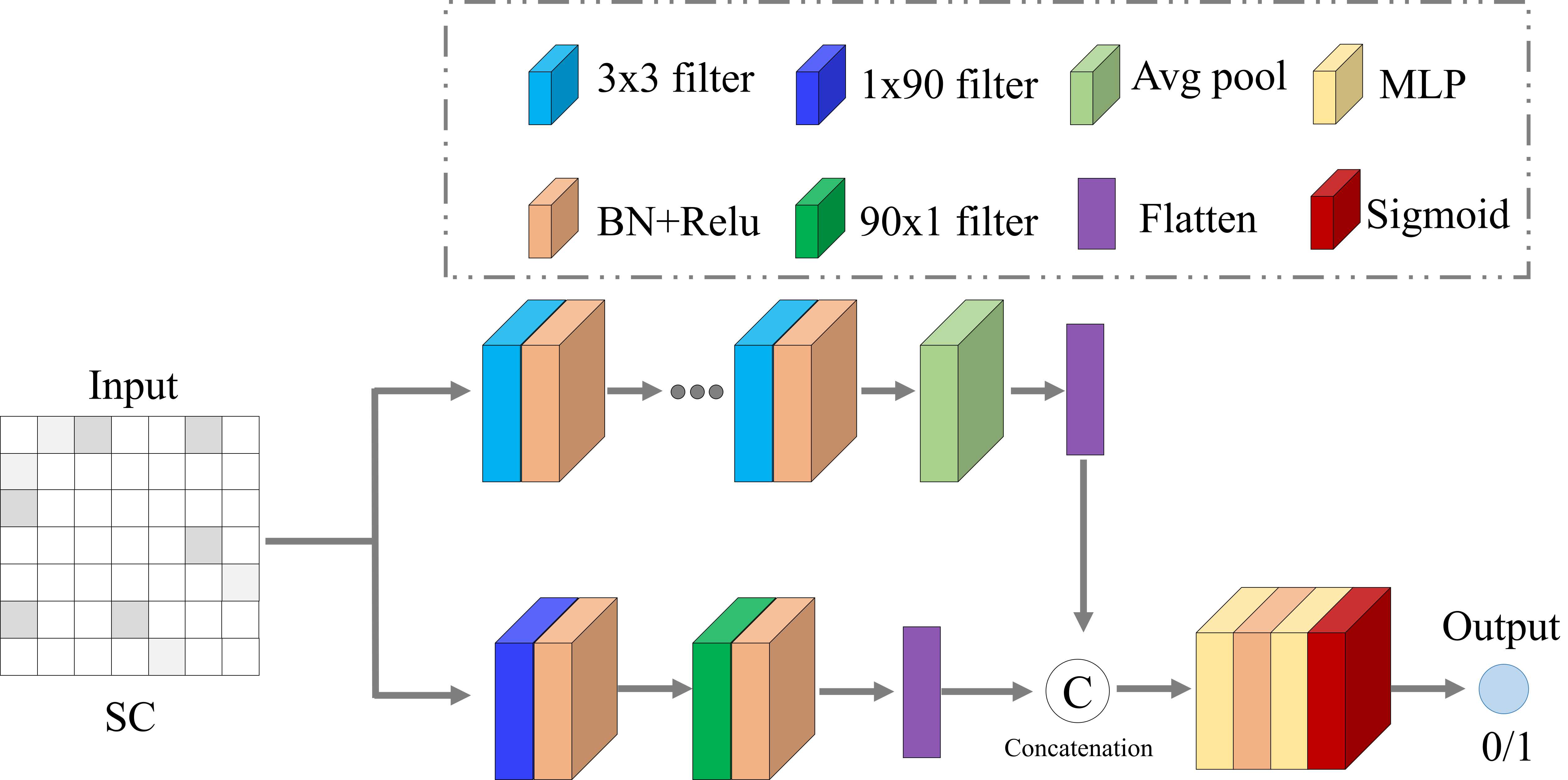}
	\caption{The network architecture of one sub-discriminator in the SFC discriminator.
	} \label{fig3}
\end{figure}

\subsection{Structural-Functional Consintency Discriminator}

The reconstructed SC and FC should also show the same distributional consistency as the empirical SC and FC. This makes the dual-channel separator robust and even diversifies the generator's abilities. The SFC discriminator contains two sub-discriminators (i.e., $D_s$ and $D_f$). Both discriminators share the same network structure. For the sake of narration, we take $D_s$ as an example to describe the detailed computing process. The input SC is passed through two branches, including local convolution (top) and global convolution (bottom). The top branch contains four $3 \times 3 $ convolution operations, one average pooling operation, and one flattened layer. The bottom branch contains two kinds of convolution filters ($1\times 90$ and $90 \times 1$) and one flattened layer. The final output is a value ranging from $0 \sim 1$, which indicates whether the input is an empirical or reconstructed SC.

\subsection{Loss Function}
Semantic feature is necessary for success in problems involving profound structure-function fusion. As a result, the proposed CT-GAN should be able to decompose a multimodal connection matrix and produce the appropriate empirical FC matrix and empirical SC matrix. The aforementioned method is accomplished using a generative adversarial technique. The network is optimized using a hybrid loss function that incorporates three types of objective losses: the adversarial loss, the classification loss, and the pair-wise connectivity reconstruction loss. This ensures the quality of the created multimodal connection. The loss functions are as follows:

\textbf{Adversarial Loss.} To make the FC' matrix and SC' matrix decoded from the multimodal connectivity matrix as similar to empirical FC matrices and SC matrices as possible, the adversarial losses are defined as follows.
\begin{equation}
	\mathcal{L}_{adv}^{\text{SC}} = \mathbb{E}_{x \sim P_{SC}}[\log D_1(\text{x})]+\mathbb{E}_{x \sim P_{DTI}}[\log (1- D_1(DS_1(G(x))))],
\end{equation}
\begin{equation}
	\mathcal{L}_{adv}^{\text{FC}} = \mathbb{E}_{x \sim P_{FC}}[\log D_2(\text{x})]+\mathbb{E}_{x \sim P_{fMRI}}[\log (1- D_2(DS_2(G(x))))],
\end{equation}
where $G$ is the designed generator, $D_1$ and $D_2$ denote the discriminators $D_s$ and $D_f$, which are the top and bottom branches in Fig.~\ref{fig2}. The distributions of the empirical SC matrix are represented by $P_{SC}$ and the empirical FC matrix by $P_{FC}$, respectively. The generator $G$ outputs the MC from fMRI and DTI. The FC matrix and SC matrix are created from the MC by the dual-channel separators $DS 1$ and $DS 2$, respectively.
Between the FC matrix decoded by $DS_1$ and empirical FC matrices, the discriminator $D_1$ tries to make a distinction. The generative adversarial strategy is that while $D$ seeks to maximize the aforementioned adversarial losses, $G$ seeks to minimize them.

\textbf{Classification Loss.} Since MC matrices can accurately predict the stages of AD, it becomes a key indicator of the effectiveness of cross-modal fusing. When the generator are optimized, the classification loss is imposed. The following provides the classification loss formula:
\begin{equation}
	\mathcal{L}_{cls} = \mathbb{E}_{x\sim P_{DTI},y\sim P_{fMRI}}[-\log p(Y|C(G(x,y)))],
\end{equation}
where normal controls (NC), early mild cognitive impairment (EMCI), late mild cognitive impairment (LMCI), and Alzheimer's disease (AD) are phases of AD (AD). The probability that the subject is currently in stage $Y$ is indicated by the classifier's output $p(Y|C(G(x,y)))$. The generator is utilized to combine features from fMRI and DTI that carry more disease information by lowering classification loss. Multimodal connectivity matrices allow the classifier to achieve the best prediction accuracy.

\textbf{Pair-wise Connectivity Reconstruction Loss.} The generator $G$ is subjected to an extra topological constraint using the $L1$ pair-wise connectivity reconstruction loss. This implies that the discriminators $D_1$ and $D_2$ must be deceived by the separators $DS_1$ and $DS_2$. The overall pair-wise connection gap between empirical FC/SC matrices and FC/SC matrices decoded by $DS_1/DS_2$ must also be reduced. These are the formalized $L1$ pair-wise reconstruction losses::
\begin{equation}
	\mathcal{L}_{pcr}^{\text{FC}} = \mathbb{E}_{x \sim P_{DTI}}\|\text{FC} - DS_1(G(x))\|_1,
\end{equation}
\begin{equation}
	\quad \mathcal{L}_{pcr}^{\text{SC}} = \mathbb{E}_{x \sim P_{fMRI}}\|\text{SC} - DS_2(G(x))\|_1.
\end{equation}

\begin{table}[h]
	\caption{Data information in this study}
	\label{tab1}
	\centering
	\resizebox{\columnwidth}{!}{
		\begin{tabular}{c c c c c c}
			\toprule[2pt]
			\textbf{Group} & \textbf{NC}(84) & \textbf{EMCI}(80) & \textbf{LMCI}(41) & \textbf{AD}(63) \\
			\midrule[1pt]
			Male/Female & 38M/46F & 48M/32F & 20M/21F  &  39M/24F  \\
			Age(mean $\pm$ SD)  & 74.0 $\pm$ 5.9 & 75.8 $\pm$ 6.1 & 74.9 $\pm$ 5.3 &  75.3 $\pm$ 5.5   \\
			\bottomrule[2pt]
	\end{tabular}}
\end{table}

\section{Experiments}
\label{s3}

\subsection{Preprocessing and Settings}

The ADNI (Alzheimer's Disease Neuroimaging Initiative) public dataset is used to test our CT-GAN model. Table~\ref{tab1} contains full information about the 268 patients whose data we used in this study. Each patient was scanned with both DTI and fMRI. The preprocessing procedure makes use of the AAL 90 atlas. Using the DPARSF toolkit, the top 20 volumes are eliminated, followed by head motion correction, band-pass filtering, Gaussian smoothing, and extracting the time series of all voxels. By following fiber bundles between ROIs, the structural connection is computed. The requirements are configured in PANDA as the fiber tracking halting conditions: a crossing angle of greater than 45 degrees between two traveling directions.

\begin{figure}[htbp]
	\centering
	\includegraphics[width=\columnwidth]{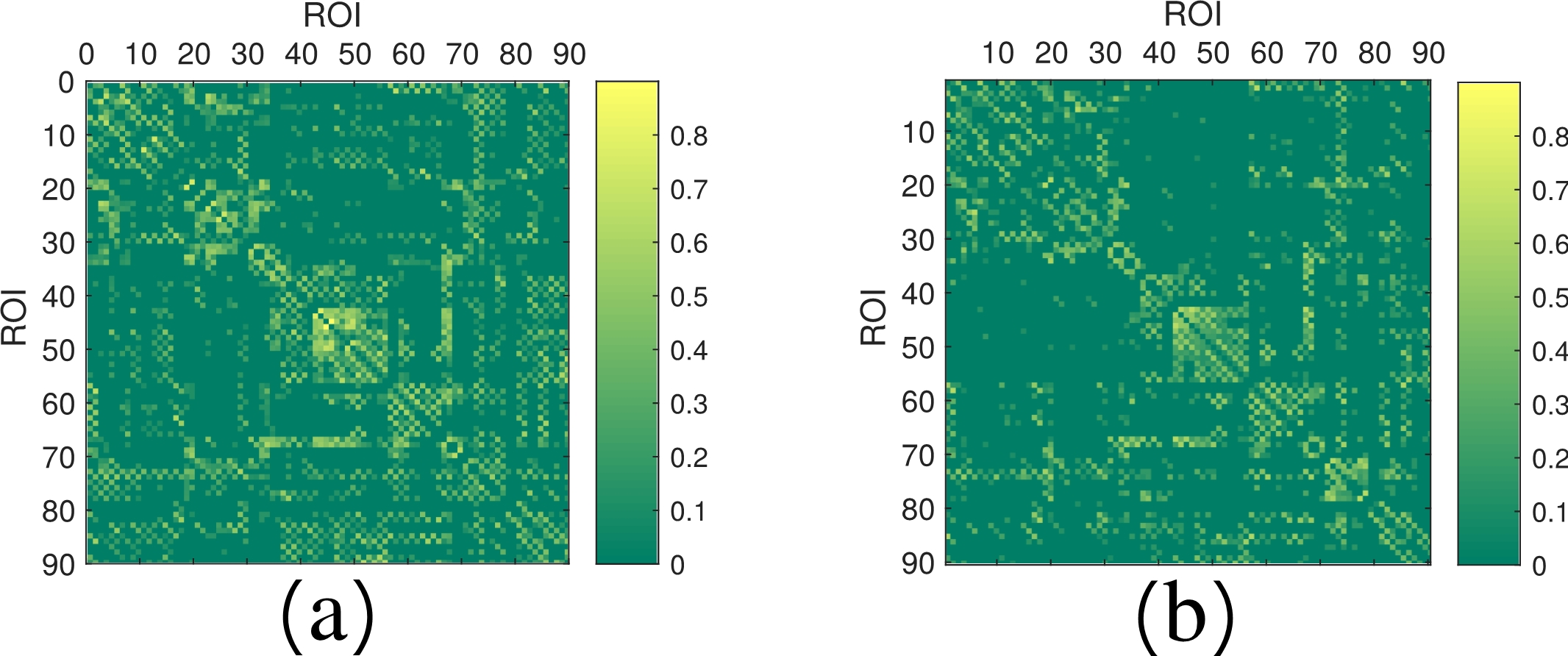}
	\caption{Examples of two multimodal connectivity matrices at different stages of cognitive disease (a) NC; (b) AD.} \label{fig4}
\end{figure}

The predictor is implemented by the row-based filters in the work \cite{kawahara2017brainnetcnn}. The embedding dimension in the generator $G$ is set at 128. $L=5$ layers of transformer are utilized to fuse structural and functional embeddings. The heads in the transformer block is 8. The end-to-end training is done on the CT-GAN model. The model's parameters will be updated during the training process using the Adam algorithm. The learning rate is set to 0.001. The weight decay is set to 0.01. The four widely used metrics accuracy (ACC), sensitivity (SEN), specificity (SPE), and area under the receiver operating characteristic curve (AUC) make up the evaluation criteria.

\begin{figure}[htbp]
	\centering
	\includegraphics[width=\columnwidth]{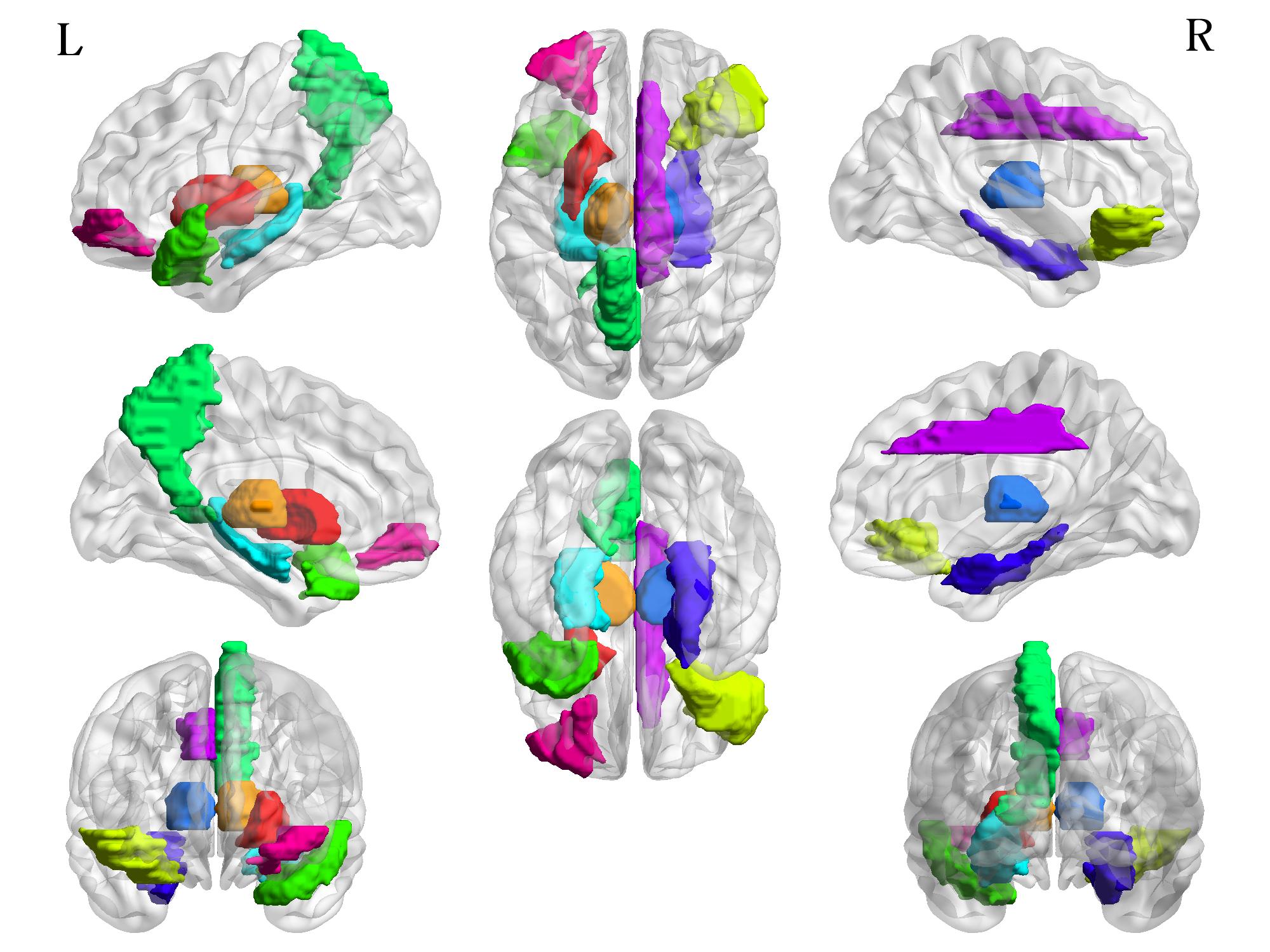}
	\caption{The ten most important brain regions between NC and EMCI groups.
	} \label{fig5}
\end{figure}

\begin{table*}[htbp]
	\renewcommand\arraystretch{1.4}
	\setlength{\abovecaptionskip}{0pt}%
	\setlength{\belowcaptionskip}{10pt}%
	\caption{Prediction of performance under different models and classifiers by fusing fMRI and DTI(\%).}
	\label{tab2}
	\resizebox{\textwidth}{20mm}{
		\begin{tabular}{c|c|cccc|cccc|cccc}
			\hline
			\multirow{2}{*}{Models} & \multirow{2}{*}{Classifiers}     & \multicolumn{4}{c|}{NC vs. EMCI}                                                                               & \multicolumn{4}{c|}{EMCI vs. LMCI}                                                                             & \multicolumn{4}{c}{LMCI vs. AD}                                                                               \\ \cline{3-14}
			&                                  & ACC                       & SEN                       & SPE                       & AUC                        & ACC                       & SEN                       & SPE                       & AUC                        & ACC                       & SEN                       & SPE                       & AUC                       \\ \hline
			MMEGCN                  &                                  & 87.20                     & 91.25                     & 83.33                     & 89.35                      & 89.26                     & 87.80                     & 90.00                     & 90.70                      & 92.31                     & 92.06                     & 92.68                     & 93.46                     \\
			GBDM                    & GCN                              & 87.80                     & 85.00                     & 90.48                     & 90.95                      & 88.43                     & 87.80                     & 88.75                     & 89.45                      & 93.27                     & 93.65                     & 92.68                     & 94.39                     \\
			Ours                    &                                  & {\bfseries 89.63}                     & 90.00                     & 89.29                     & {\bfseries 93.44}                      & {\bfseries 93.39}                     & {\bfseries 92.68}                     & {\bfseries 93.75}                     & {\bfseries 94.51}                      & {\bfseries 94.23}                     & 93.65                     & {\bfseries 95.12}                     & {\bfseries 95.20}                     \\ \hline
			MMEGCN                  & \multicolumn{1}{l|}{}            & \multicolumn{1}{l}{88.41} & \multicolumn{1}{l}{87.50} & \multicolumn{1}{l}{89.29} & \multicolumn{1}{l|}{90.71} & \multicolumn{1}{l}{88.43} & \multicolumn{1}{l}{90.24} & \multicolumn{1}{l}{87.50} & \multicolumn{1}{l|}{88.17} & \multicolumn{1}{l}{91.35} & \multicolumn{1}{l}{90.48} & \multicolumn{1}{l}{92.68} & \multicolumn{1}{l}{89.47} \\
			GBDM                    & \multicolumn{1}{l|}{Brainnetcnn} & \multicolumn{1}{l}{88.41} & \multicolumn{1}{l}{90.00} & \multicolumn{1}{l}{86.90} & \multicolumn{1}{l|}{90.31} & \multicolumn{1}{l}{91.74} & \multicolumn{1}{l}{90.24} & \multicolumn{1}{l}{92.50} & \multicolumn{1}{l|}{89.42} & \multicolumn{1}{l}{93.27} & \multicolumn{1}{l}{92.06} & \multicolumn{1}{l}{95.12} & \multicolumn{1}{l}{91.99} \\
			Ours                    & \multicolumn{1}{l|}{}            & \multicolumn{1}{l}{{\bfseries 90.24}} & \multicolumn{1}{l}{90.00} & \multicolumn{1}{l}{{\bfseries 90.48}} & \multicolumn{1}{l|}{{\bfseries 93.26}} & \multicolumn{1}{l}{{\bfseries 92.56}} & \multicolumn{1}{l}{90.24} & \multicolumn{1}{l}{{\bfseries 93.75}} & \multicolumn{1}{l|}{{\bfseries 93.93}} & \multicolumn{1}{l}{{\bfseries 95.19}} & \multicolumn{1}{l}{{\bfseries 95.24}} & \multicolumn{1}{l}{95.12} & \multicolumn{1}{l}{{\bfseries 94.27}} \\ \hline
	\end{tabular}}
\end{table*}

\subsection{Prediction Results}
To demonstrate the generation effect of the proposed model, Fig.~\ref{fig4} qualitatively depicts four examples of the generated MC at different stages. Even though the four MCs show the same global connectivity patterns, different connectivity characteristics can be seen in the local area. The MC at the AD stage has the sparsest connectivity features.

\begin{figure}[htbp]
	\centering
	\includegraphics[width=\columnwidth]{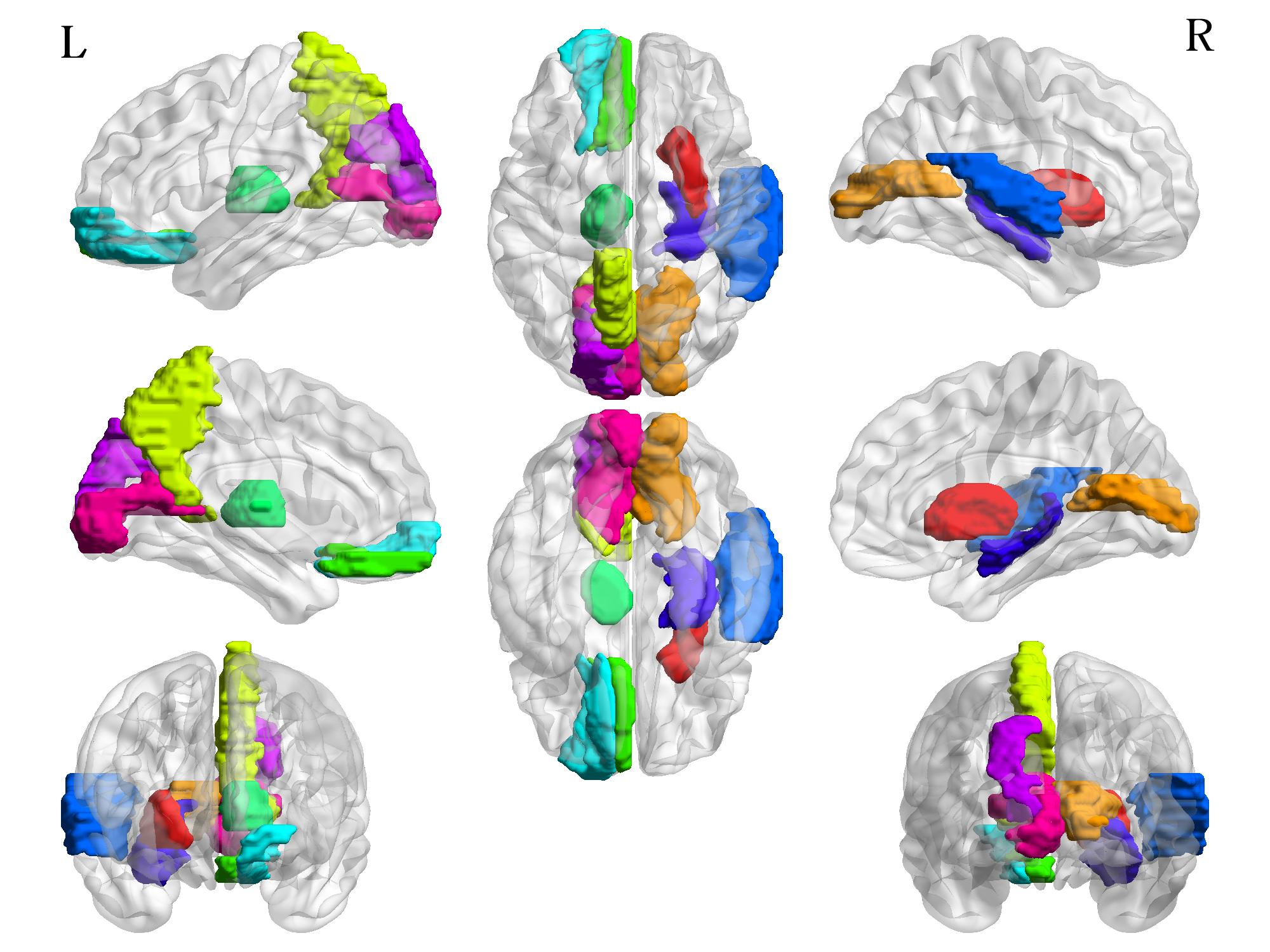}
	\caption{The ten most important brain regions between EMCI and LMCI groups.
	} \label{fig6}
\end{figure}

\begin{figure}[htbp]
	\centering
	\includegraphics[width=\columnwidth]{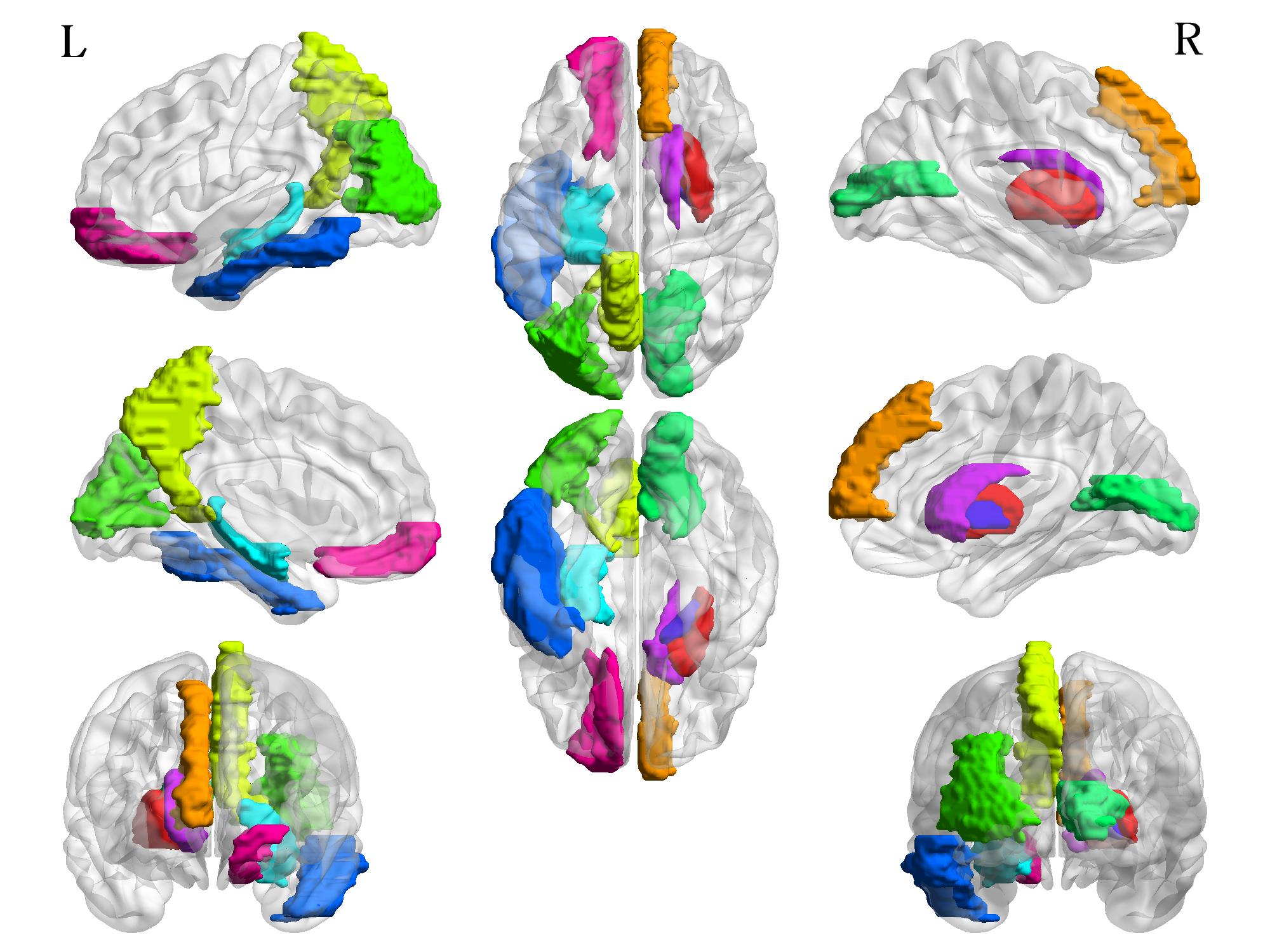}
	\caption{The ten most important brain regions between LMCI and AD.
	} \label{fig7}
\end{figure}

\begin{figure*}[htbp]
	\centering
	\includegraphics[width=0.9\textwidth]{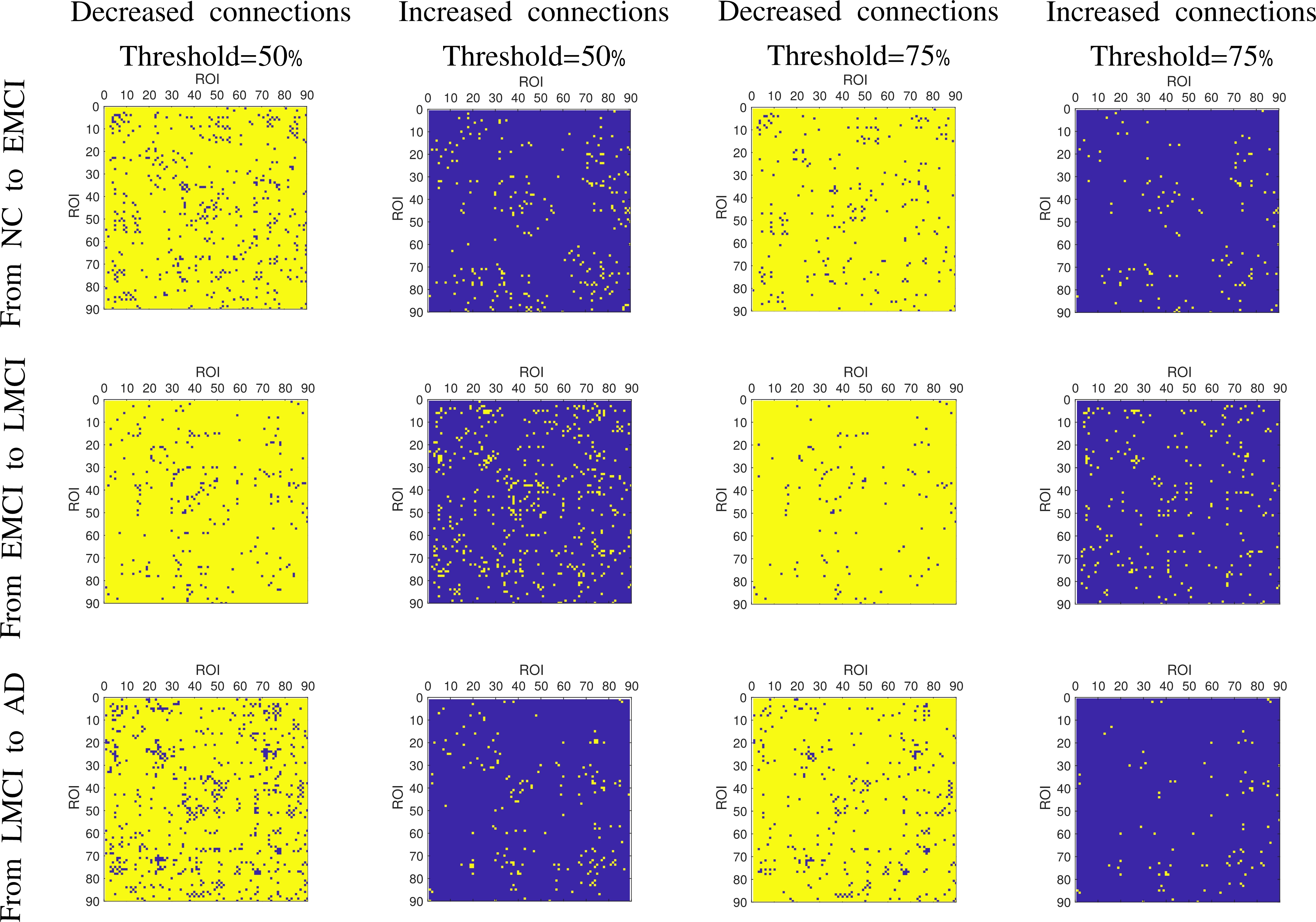}
	\caption{The altered connectivity between two MC groups. The first and third columns are the decreased connectivity matrices, with the threshold values set at 50\% and 75\% respectively. The second and fourth columns are the increased connectivity matrices with the threshold values at 50\% and 75\% respectively.} \label{fig8}
\end{figure*}

To conduct a quantitative analysis of the proposed model's classification, we conducted three binary classification tasks (i.e., NC vs. EMCI, EMCI vs. LMCI, and LMCI vs. AD). Each classification task is operated with the five-fold cross-validation strategy. To evaluate how well various fMRI-DTI fusion models can capture characteristics associated with AD, two competing models and two classifiers are introduced in our experiments for comparison. Specifically, the multi-modal enhanced graph convolutional network (MMEGCN) \cite{cls1} and the graph-based deep model (GBDM)\cite{zhang2021deep} output a combined brain network by inputting fMRI and DTI. After the combined brain network has been generated, we adopt two classifiers (GCN\cite{kipf2016semi} and Brainnetcnn\cite{kawahara2017brainnetcnn}) to evaluate the classification performance of the generated brain networks. Table~\ref{tab2} shows the detailed prediction performance between different competing models. Under different classification tasks, our model achieves superior results to others in terms of different classifiers. Both classifiers have similar classification performances. The best classification results for NC vs. EMCI are ACC value of 90.24\%, SEN value of 90.00\%, SPE value of 90.48\%, and AUC value of 93.26\%; the EMCI vs. LMCI task achieves the best ACC value of 93.39\%, SEN value of 92.68\%, SPE value of 93.75\%, and AUC value of 94.51\%; the best results for LMCI vs. AD are ACC value of 95.19\%, SEN value of 95.24\%, SPE value of 95.12\%, and AUC value of 94.27\%.Overall, the results of the experiments demonstrate that the proposed CT-GAN has the benefit of being more accurate than previous multimodal fusion models in predicting the phases of AD.

\begin{figure*}[htbp]
	\centering
	\includegraphics[width=0.99\textwidth]{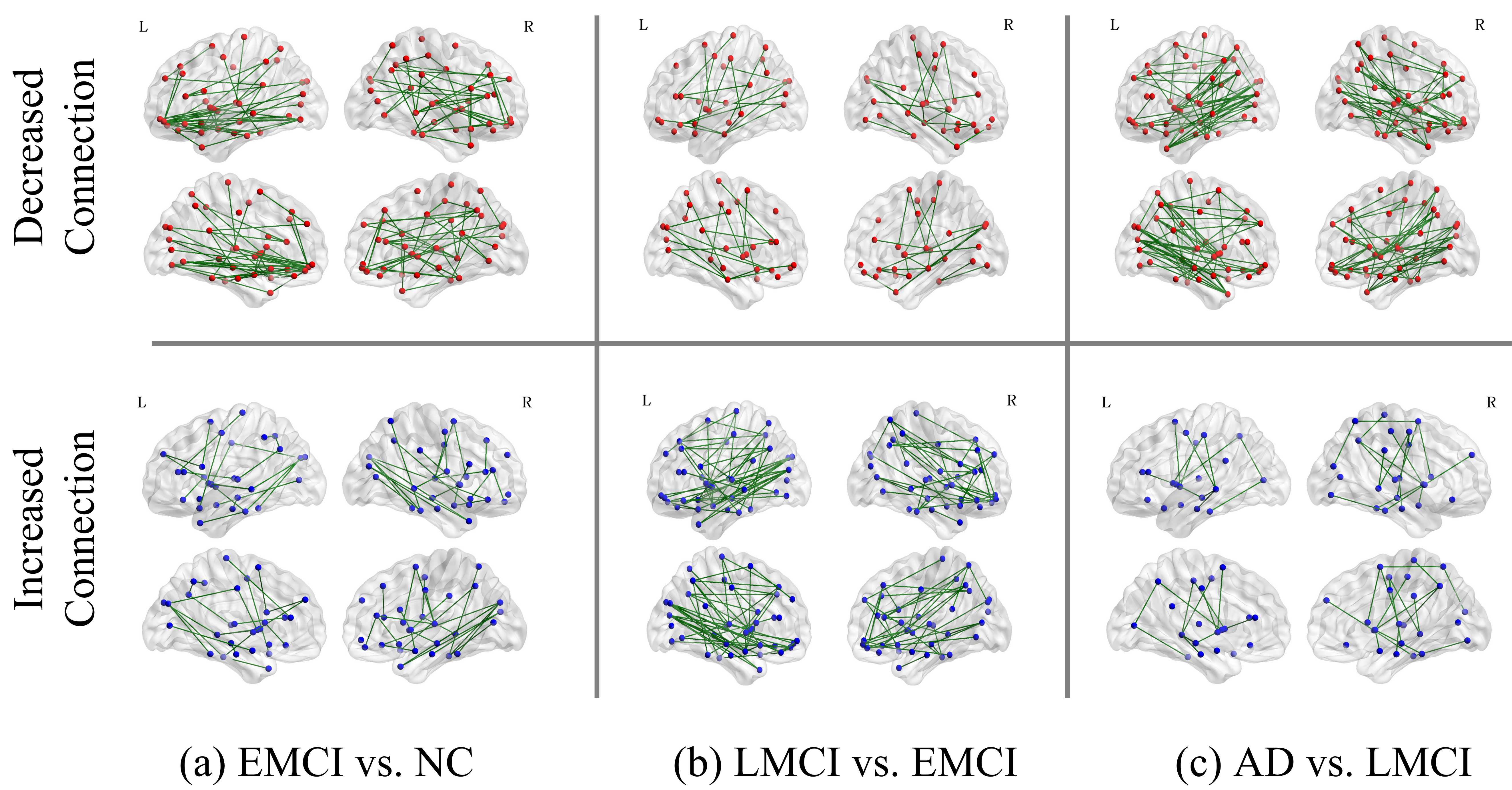}
	\caption{The altered connectivities with the threshold value at 75\% quantile for the three scenarios.
	} \label{fig9}
\end{figure*}

To evaluate the AD-related ROIs in the classification tasks, we utilized the LOOCV method \cite{mci3} to compute the important score for each ROI during the ACC prediction. The important ROIs are displayed by the BrainNetviewer tool\cite{xia2013brainnet}. As shown in Fig.~\ref{fig5}, the ten important ROIs for NC vs. EMCI are the left lenticular nucleus putamen, left Thalamus, right inferior frontal gyrus orbital part, left Temporal pole superior temporal gyrus, left Precuneus, left Hippocampus, right Thalamus, right Parahippocampal gyrus, right Median cingulate and paracingulate gyri, and left Middle frontal gyrus orbital part. For EMCI vs. LMCI, the identified 10 important ROIs in Fig.~\ref{fig6} are the right Lenticular nucleus putamen, the right Calcarine fissure and surrounding cortex, the left Precuneus, the left Gyrus rectus, the left Thalamus, the left Superior frontal gyrus orbital part, the right Superior temporal gyrus, the right Hippocampus, the left Superior occipital gyrus, and the left Calcarine fissure and surrounding cortex. In Fig.~\ref{fig7}, the important ROIs between the LMCI and AD groups are the following: right Lenticular nucleus putamen, right Superior frontal gyrus medial, left Precuneus, left Middle occipital gyrus, right Calcarine fissure and surrounding cortex, left Hippocampus, left Inferior temporal gyrus, right Lenticular nucleus pallidum, right Caudate nucleus, left Superior frontal gyrus orbital part. The identified important ROIs are found to be related with alzheimer's disease and is partly overlapped with previous studies\cite{li2019multimodal,lei2021auto,lei2023multi}.

\begin{figure}[htbp]
	\centering
	\includegraphics[width=0.9\columnwidth]{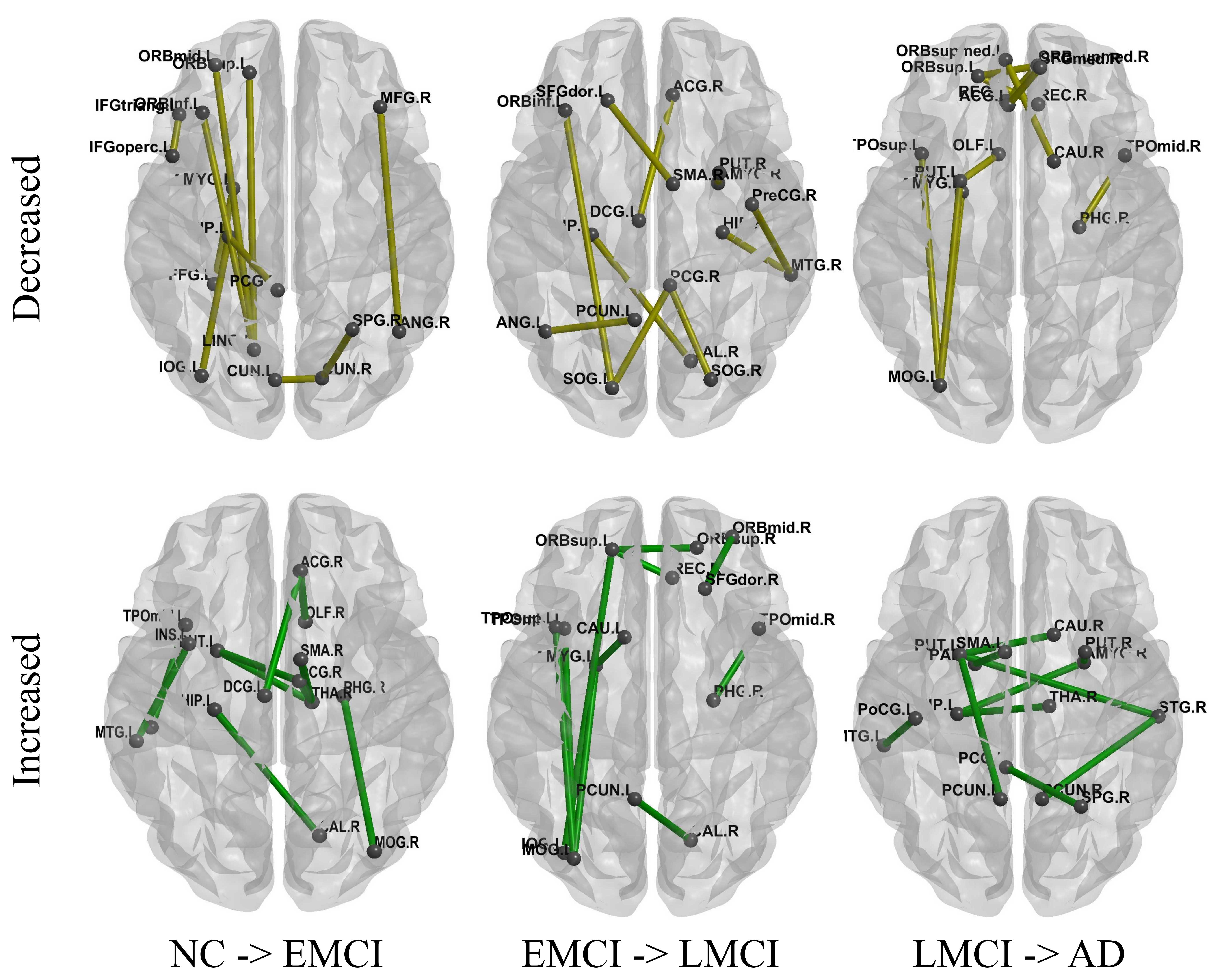}
	\caption{The spatial view of the top 10 decreased and increased connections for the three scenarios.
	} \label{fig10}
\end{figure}

\subsection{Connectivity Analysis}


To analyze the important connections associated with AD, we first compute the averaged MC matrix for each disease group (i.e., NC, EMCI, LMCI, and AD) and then evaluate the difference matrix between adjacent groups. The positive value in the difference matrix means the increased connections and the negative value represents the decreased connections. The averaged multimodal connectivity matrices at different stages of AD disease can be obtained by applying the trained generator on DTI and fMRI. The visualization of averaged multimodal connectivity matrices and the change in connectivity with various thresholds are shown in Fig.~\ref{fig8}. The three rows correspond to the altered connections from NC to EMCI, from EMCI to LMCI, and from LMCI to AD, respectively. The values between $-0.1 \sim 0.1$ are ignored during the analysis. Two threshold values are set for viewing the important connections. The first threshold is 50\% quantile values, which are estimated from the positive and negative connectivities. The same operation is implemented on the second 75\% threshold value. The more important connections with the 75\% threshold value are shown in Fig.~\ref{fig9}. It can be seen that the decreased connections are greater than the increased connections at the stages of EMCI and AD, while the phenomenon is reversed at the LMCI stage.

\begin{figure*}[htbp]
	\centering
	\includegraphics[width=0.9\textwidth]{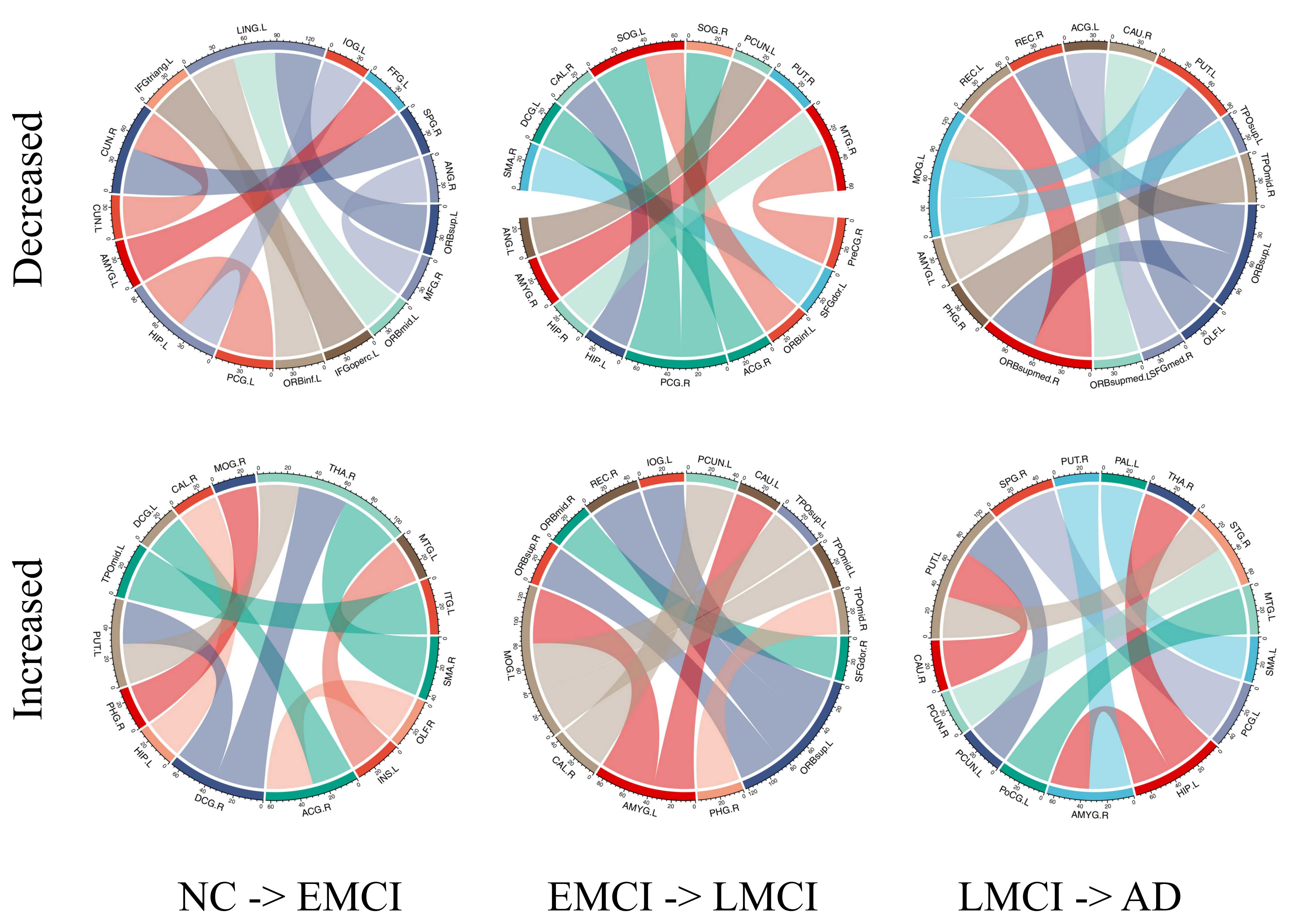}
	\caption{A view of the circos plot for the corresponding top 10 decreased and increased connections.
	} \label{fig11}
\end{figure*}

\begin{figure*}[htbp]
	\centering
	\includegraphics[width=0.8\textwidth]{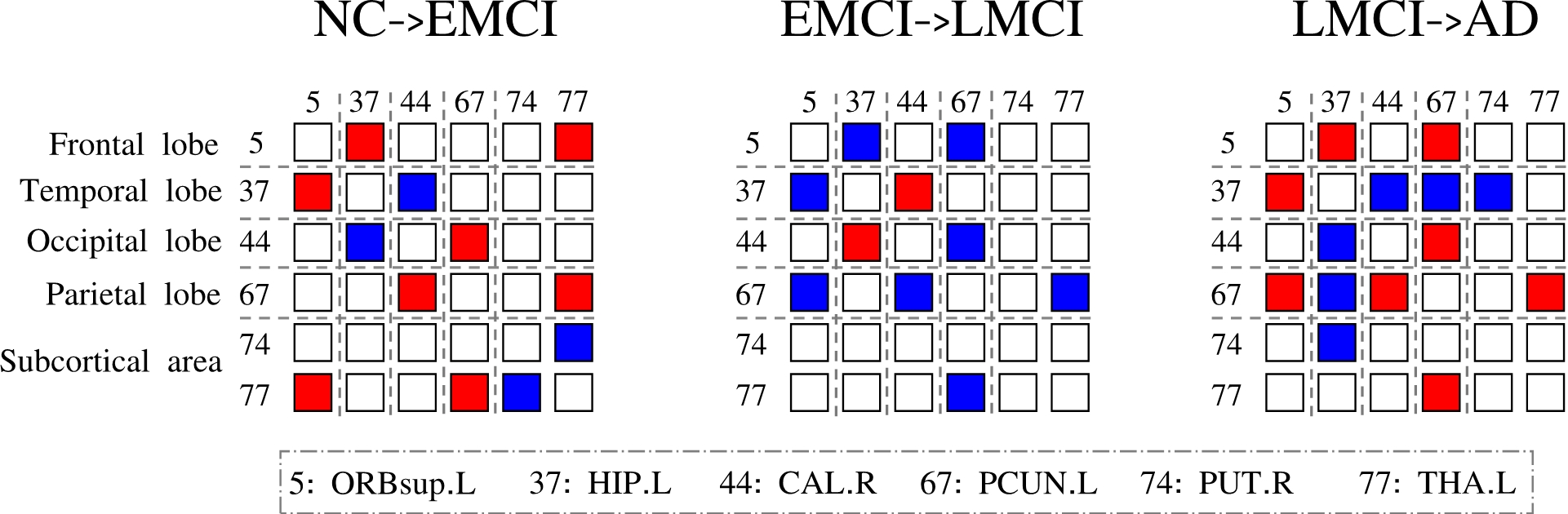}
	\caption{The altered multimodal connectivities associated with the overlapping six ROIs in the prediction results. The index indicates the corresponding ROI in the AAL90 atlas. The red color represents decreased connections; the blue color represents increased connections. The gray dotted lines divide the six ROIs into five brain lobes.
	} \label{fig12}
\end{figure*}

To evaluate the most important connections for different stages of AD, we sort the altered connection strength and find the top 10 connections for both increased and decreased situations. The results are shown in  Fig.~\ref{fig10} and Fig.~\ref{fig11}. For the NC vs. EMCI, the increased connections are SMA.R - THA.R, TPOmid.L - ITG.L, DCG.R - THA.R, OLF.R - ACG.R, INS.L - MTG.L, DCG.R - PUT.L, ACG.R - DCG.L, PUT.L - THA.R, PHG.R - MOG.R, HIP.L - CAL.R; the decreased connections are PCG.L - HIP.L, IFGoperc.L - IFGtriang.L, ORBsup.L - LING.L, CUN.R - SPG.R, ORBinf.L - LING.L, AMYG.L - FFG.L, MFG.R - ANG.R, ORBmid.L - LING.L, HIP.L - IOG.L, CUN.L - CUN.R. As the stage changes from EMCI to LMCI, the increased connectivities are the following: AMYG.L - MOG.L, ORBsup.L - REC.R, CAL.R - PCUN.L, MOG.L - TPOsup.L, MOG.L - TPOmid.L, PHG.R - TPOmid.R, ORBsup.L - ORBsup.R, ORBsup.L - IOG.L, SFGdor.R - ORBmid.R, AMYG.L - CAU.L; and the decreased connectivities are the following: PCG.R - SOG.L, PreCG.R - MTG.R, AMYG.R - PUT.R, SFGdor.L - SMA.R, PCG.R - SOG.R, ACG.R - DCG.L, ORBinf.L - SOG.L, HIP.L - CAL.R, HIP.R - MTG.R, ANG.L - PCUN.L. The top 10 connections from LMCI to AD are increased at PCG.L - SPG.R, HIP.L - THA.R, CAU.R - PUT.L, PoCG.L - MTG.L, PCUN.L - PUT.L, SMA.L - PAL.L, AMYG.R - PUT.R, PUT.L - STG.R, PCUN.R - STG.R, HIP.L - AMYG.R; and decreased at ORBsupmed. R - REC.L, ORBsup.L - REC.R, ORBsup.L - ORBsupmed.R, PHG.R - TPOmid.R, AMYG.L - MOG.L, ORBsupmed.L - CAU.R, OLF.L - PUT.L, SFGmed.R - ACG.L, MOG.L - PUT.L, MOG.L - TPOsup.L.

To show the significant important ROI-related connections, we counted the overlapping brain regions of the three top 10 ROIs in the prediction results and found six frequent ROIs. These ROIs include the following: left Superior frontal gyrus orbital part, left Hippocampus, right Calcarine fissure and surrounding cortex, left Precuneus, right Lenticular nucleus putamen, left Thalamus. Based on the increased and decreased connectivities, the important ROI-related connections are shown in Fig.~\ref{fig12}. Even though these connections are not listed in the top 10 altered connections, they reflect the changing characteristics at the different stages of AD. For example, the connection ORBsup.L-HIP.L first drops in strength at EMCI, then gains some strength at LMCI, and finally loses strength again at the AD stage. These findings can be explained by the compensation mechanism \cite{montembeault2016altered,berron2020medial}.

\begin{figure}[h]
	\centering
	\includegraphics[width=\columnwidth]{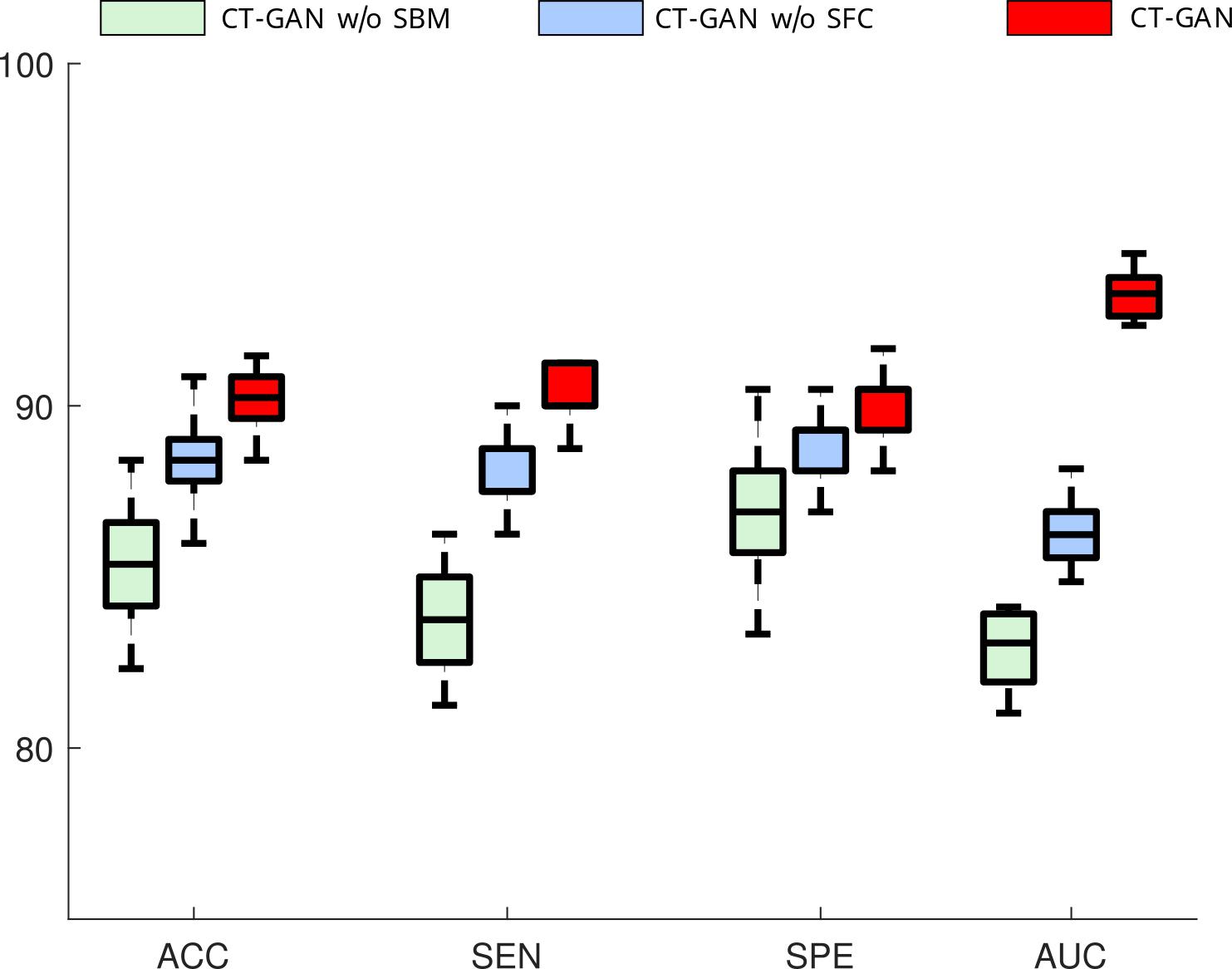}
	\caption{Influence of different modules in CT-GAN on the prediction performance.
	} \label{fig13}
\end{figure}

\subsection{Ablation Studies}
The proposed CT-GAN can generate discriminative MCs for early AD analysis. The prediction precision and altered connectivity evaluation proved the effectiveness of the model. To study the impact of different modules in the model on the prediction results, we (1) replace the SBM module with a traditional multi-head self-attention mechanism; and (2) remove the SFC discriminator. As shown in Fig.~\ref{fig13}, the removal of the SBM mechanism significantly lowers the prediction performance. As well, the SFC discriminator affects the classification accuracy to some extent.

\section{Conclusion}
\label{s4}

In this paper, we propose a novel CT-GAN model to fuse fMRI and DTI and generate multimodal connectivity from fMRI and DTI in an efficient end-to-end manner. The key idea of this work is that mutual conversion between structural and functional information is accomplished using a cross-modal swapping bi-attention mechanism. Therefore, the proposed model can gradually and effectively extract complementary information between modalities. The results of the experiments demonstrated that the multimodal connectivity generated by our model is more accurate than other multimodal fusion models in terms of classification performance. Furthermore, some AD-related connectomes and brain regions are identified by analyzing the generated multimodal connectivity matrices. These connectomes partially agree with the clinical investigations on AD, which indicates that the proposed model can provide new insights for detecting AD-related abnormal connectivities. In the future, we will extend the CT-GAN to other neurodegenerative diseases for evaluation and analysis.


\section*{Acknowledgement}
This research is supported by the National Natural Science Foundations of China under Grant No. 61872351, Shenzhen Key Basic Research Project under Grant No.JCYJ20180507182506416, ADNI (National Institutes of Health Grant U01 AG024904), DOD ADNI (Department of Defense award number W81XWH-12-2-0012) and HKRGC Grant Numbers: GRF 12200317, 12300218, 12300519 and 17201020.

\end{document}